\documentclass[preprint,showpacs,preprintnumbers,amsmath,amssymb]{revtex4}
\usepackage{graphicx,epsfig}

\usepackage{graphicx}
\usepackage{dcolumn}
\usepackage{bm}

\begin{document}


\title{Transmission resonances and supercritical states in a one dimensional
cusp potential}

\author{V\'{\i}ctor M. Villalba\footnote{Alexander von Humboldt Fellow}}
\email{villalba@ivic.ve}
\affiliation{Centro de F\'{\i}sica IVIC Apdo 21827, Caracas 1020A, Venezuela
}%

\author{Walter Greiner}%
\affiliation{Institut f\"{u}r Theoretische Physik.
Universit\"{a}t Frankfurt \\  D-60054 Frankfurt am Main, Germany
}

\date{\today}

\begin{abstract}
We solve the two-component Dirac equation in the presence of a spatially one 
dimensional symmetric cusp potential.  We compute the scattering and bound 
states solutions and we derive the conditions for transmission resonances 
as well as for supercriticality. 
\end{abstract}


\maketitle

\section{Introduction}

The study of scattering and bound states of nonrelativistic particles in
the Schr\"{o}dinger equation framework is a well known and understood
problem \cite{Newton}. The scattering of low momentum particles by one-dimensional
potentials that are well behaved at infinity has zero transmission coefficient at
zero energy, and the reflection coefficient is unity unless the potential supports
a half bound state (zero energy resonance). In that case there is no
reflection and the transmission coefficient is unity provided that the
potential is symmetric. The pioneering works on the computation of
transmission and reflection coefficients in one-dimensional potentials go
back to Bohm \cite{Bohm} who coined the definition of transmission
resonance to scattering states with transmission coefficient equal to
unity and vanishing reflection. Recently \cite{Senn,Sassoli}, these results
have been generalized to asymmetric potentials.

The generalization of the concept of transition resonances to relativistic
quantum mechanics presents some subtleties related to the
Dirac equation \cite{Kennedy2,Dombey}. For a free Dirac particle there are
positive energy states $(E>m)$ as well as negative energy states $(E<-m)$.
In the presence of an external potential, half bound states occur for $E=-m$
and $E=m$, in contrast to the non-relativistic case where these only exist
at $E=0.$ In the Dirac equation framework one should talk about zero
momentum resonances rather than zero energy resonances.

The study of scattering and bound states in the presence of strong
electromagnetic fields in relativistic quantum mechanics  is also a rather
old problem. Soon after the formulation of the Dirac equation \cite{Dirac},
Klein \cite{Klein} found that for very high potential barriers an unusually
large number of electrons penetrate into the wall. Moreover, those electrons
have negative
energy. Sauter \cite{Sauter} found the same results in the more general case
where the barrier smoothly grows. Pair creation for the Klein
paradox is stimulated by the incoming electron beam. When the depth of a
square well exceeds $2mc^{2}$, the energy of the lowest bound state becomes
equal to $-mc^{2}$ , and the bound level joins the states of the Dirac sea.
The bound states dive into the energy continuum \cite{Greiner}.

The Klein paradox in a one-dimensional Saxon-Woods potential was studied by
Dosch {\it et al} \cite{Dosch}, showing that in this case transmission resonances exist.
Kennedy \cite{Kennedy} discussed supercriticality in
this potential, demonstrating that, as in the square potential case, 
the one-dimensional Woods-Saxon potential has transmission resonances of zero momentum 
when it supports a half bound state at $E=-m$.

The Saxon-Woods potential presents the particularity of being a smoothed
form of a square well; the Dirac equation presents half bound states with
the same asymptotic behavior as those obtained with a potential barrier 
\cite{Kennedy}.

\begin{figure}
\includegraphics[width=12cm]{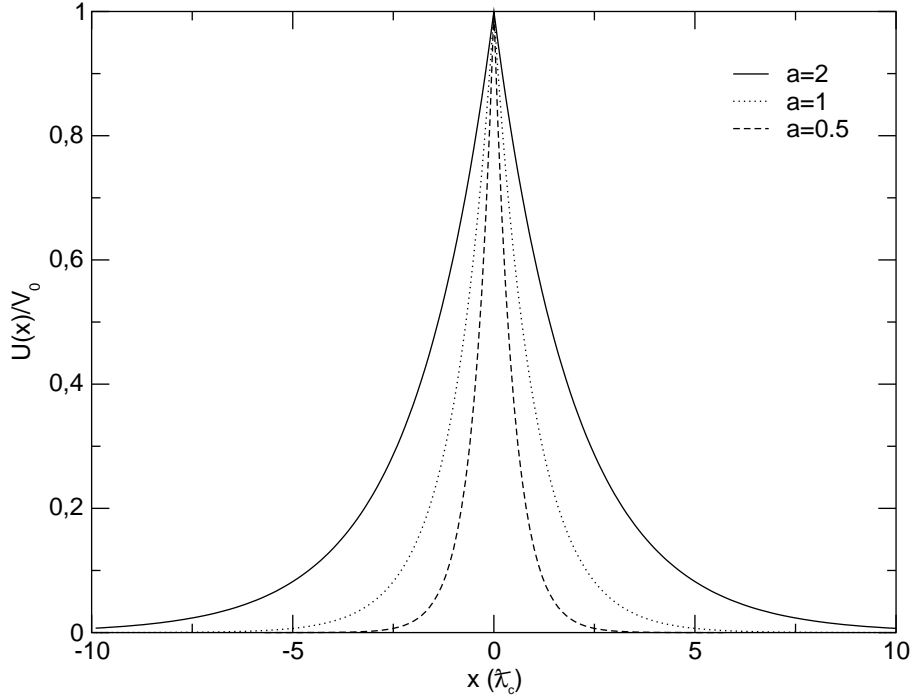}
\caption{\label{fig:Fig1} The screened potential for different values of the  
shape parameter $a (\lambdabar_{c})$. The solid line corresponds to $a=2$, 
the dotted line to $a=1$, and the dashed line to $a=0.5$}
\end{figure}

In the present article we discuss the problem of computing transmission
resonances and supercriticality of the one-dimensional symmetric screened
potential (Fig \ref{fig:Fig1}): 
\begin{equation}
U(x)=V_{0}\exp (-\left| x\right| /a).  \label{A1}
\end{equation}

The potential (\ref{A1}) is an asymptotically vanishing potential for large
values of the \ space variable $x$. The parameter $V_{0}$ shows the height of
the barrier or the depth of the well depending on its sign. The positive
constant $a$ determines the shape of the potential. A noticeable difference
between the Woods-Saxon potential and (\ref{A1}) is that this one does not
exhibit a square barrier limit. The potential (\ref{A1}), for $V_{0}<0,$ can
be regarded as a screened one dimensional Coulomb potential \cite{Dominguezadame}.

It is the purpose of the present article to show that the potential (\ref{A1}), 
for $V_{0}>0$, supports transmission resonances. We solve the 1+1 Dirac
equation explicitly and compute the bound energy states for $U(x),$ for $%
V_{0}<0,$ and show that this potential supports supercritical states which correspond to
transmission resonances for the potential (\ref{A1}), for $V_{0}>0.$

The article is structured as follows. In Sec. II we solve the 1+1 Dirac
equation in the potential (\ref{A1}). In Sec III we compute the bound states
and the condition that the potential should satisfy in order to support
supercritical states $(E=-m)$. In Sec. IV. we compute the scattering states
and show that for $k=0$ the condition for transmission resonances
corresponds to that of supercriticality for $V_0<0$.

\section{The Dirac equation}

The Dirac equation in the presence of an external electromagnetic field $%
A_{\mu}$ has the form

\begin{equation}
\left\{ \gamma ^{\mu }(\partial _{\mu }-ieA_{\mu })+m\right\} \Psi =0,
\label{odin}
\end{equation}
where the four-vector potential (\ref{A1}) can be written in a covariant way as 
\begin{equation}
A^{\mu }=V_{0}\exp (-\left| x\right| /a)\delta _{0}^{\mu }.  \label{A}
\end{equation}
In two dimensions, the Dirac equation (\ref{odin}) in the presence of the
spatially dependent electric field (\ref{A}) reads

\begin{equation}
\left\{ \gamma ^{0}(\partial _{t}+ieV_{0}\exp (-\left| x\right| /a))+\gamma
^{1}\partial _{x}+m\right\} \Psi =0  \label{Dirac}
\end{equation}
where the $\gamma ^{\mu }$ matrices satisfy the commutation relation $%
\left\{ \gamma ^{\mu },\gamma ^{\nu }\right\} =2\eta ^{\mu \nu }$ and the
metric $\eta ^{\mu \nu }$ has the signature $(-,+).$ Here and throughout the
present article we adopt units where $\hbar =c=1.$

Since we are working in 1+1 dimensions, it is possible to choose the
following representation of the Dirac matrices 
\begin{equation}
\gamma ^{0}=i\sigma ^{2}=\left( 
\begin{array}{cc}
0 & 1 \\ 
-1 & 0
\end{array}
\right) ,\ \gamma ^{1}=\sigma ^{1}=\left( 
\begin{array}{cc}
0 & 1 \\ 
1 & 0
\end{array}
\right)  \label{rep}
\end{equation}
Taking into account that the vector potential (\ref{A}) does not depend
explicitly on time, we can separate the time dependence of the spinor $\Psi $
as follows: 
\begin{equation}
\Psi =\left( 
\begin{array}{c}
\Psi _{1}(x) \\ 
\Psi _{2}(x)
\end{array}
\right) \exp (-iEt)  \label{Psi}
\end{equation}
Substituting the gamma matrices (\ref{rep}) into the Dirac equation, (\ref
{Dirac}) we obtain the following system of coupled equations 
\begin{equation}
\left( \frac{d}{dx}-i(eV_{0}\exp (-\left| x\right| /a)-E)\right) \Psi
_{1}+m\Psi _{2}=0  \label{a1}
\end{equation}
\begin{equation}
\left( \frac{d}{dx}+i(eV_{0}\exp (-\left| x\right| /a)-E)\right) \Psi
_{2}+m\Psi _{1}=0  \label{a2}
\end{equation}
Introducing the new variable 
\begin{equation}
y=2iaeV_{0}\exp (-x/a)\ x\mathrm{>0}
\end{equation}
we reduce the system (\ref{a1})-(\ref{a2}) to

\begin{equation}
\left( y\frac{d}{dy}-\frac{y}{2}+iEa\right) \Psi _{2}=ma\Psi _{1}  \label{b1}
\end{equation}
\begin{equation}
\left( y\frac{d}{dy}+\frac{y}{2}-iEa\right) \Psi _{1}=ma\Psi _{2}  \label{b2}
\end{equation}
The regular solutions of the system of equations (\ref{b1})-(\ref{b2}),can
be expressed in terms of the Whittaker function $M_{k,\mu }(y)$ \ \cite
{Abramowitz} as follows 
\begin{equation}
\Psi _{2}=b_{1}y^{-1/2}M_{k,\mu }(y)  \label{se}
\end{equation}
\begin{equation}
\Psi _{1}=\frac{1/2+\mu +k}{ma}b_{1}y^{-1/2}M_{k+1,\mu }(y)  \label{pri}
\end{equation}
where $b_{1}$ is a normalization constant, and 
\begin{equation}
k=iEa-1/2,\ \mu =i\sqrt{E^{2}-m^{2}}a  \label{ka}
\end{equation}
For negative values of $x$, we introduce the auxiliary variable $\bar{y}$ 
\begin{equation}
\bar{y}=2aieV_{0}\exp (x/a)\ x\mathrm{<0.}  \label{var}
\end{equation}
In terms of the new variable $\bar{y},$ the system of equations (\ref{a1})-(%
\ref{a2}) takes the form 
\begin{equation}
\left( \bar{y}\frac{d}{d\bar{y}}+\frac{\bar{y}}{2}-iEa\right) \Psi
_{2}+ma\Psi _{1}=0  \label{c1}
\end{equation}
\begin{equation}
\left( \bar{y}\frac{d}{d\bar{y}}-\frac{\bar{y}}{2}+iEa\right) \Psi
_{1}+ma\Psi _{2}=0  \label{c2}
\end{equation}
and its solution can be written as 
\begin{equation}
\Psi _{1}=c_{1}\bar{y}^{-1/2}M_{k,\mu }(\bar{y})  \label{se2}
\end{equation}
\begin{equation}
\Psi _{2}=-\frac{1/2+\mu +k}{ma}c_{1}\bar{y}^{-1/2}M_{k+1,\mu }(\bar{y})
\label{pri2}
\end{equation}
where $c_{1}$ is a normalization constant

\section{Bound states}

The potential (\ref{A}) supports bound states provided that $m^{2}-E^{2}>0.$
In this case one has that $\mu $ can be written as 
\begin{equation}
\mu =a\sqrt{m^{2}-E^{2}}
\end{equation}
and the condition for existence of bound energy states can be obtained after
imposing the normalizability of the wave function  as well as the
continuity of  $\Psi $ at $x=0.$ Noticing the normalizability of the
spinor $\Psi $ upon the scalar product 
\begin{equation}
\left\langle \Psi ,\Phi \right\rangle =\int \bar{\Psi}\gamma ^{0}\Phi dx
\end{equation}
imposes that the spinor components $\Psi _{1}$ and $\Psi _{2}$ should vanish
as $\left| x\right| \rightarrow \infty $. This condition is equivalent to
imposing the regularity of the spinor at $y=0,$ and $\bar{y}=0.$ The solutions
(\ref{se})-(\ref{pri}), and (\ref{se2})-(\ref{pri2}) satisfy this boundary
condition. The energy levels can be computed by imposing the continuity of
the spinor $\Psi $ at $x=0.$ 
\begin{equation}
c_{1}M_{k,\mu }(2ieaV_0)=b_{1}\frac{1/2+\mu +k}{ma}M_{k+1,\mu }(2ieaV_0)
\label{e1}
\end{equation}
\begin{equation}
b_{1}M_{k,\mu }(2ieaV_0)=-c_{1}\frac{1/2+\mu +k}{ma}M_{k+1,\mu }(2ieaV_0)
\label{e2}
\end{equation}
The existence of non-trivial $c_{1}$ and $b_{1}$ satisfying the above system
(\ref{e1})-(\ref{e2}) is equivalent to imposing the condition 
\begin{equation}
\left( \frac{1}{2}+\mu
+k\right) ^{2}M_{k+1,\mu }^{2}(2ieaV_{0})+m^{2}a^{2}M_{k,\mu
}^{2}(2ieaV_{0})=0.
\label{energy} 
\end{equation}
The expression (\ref{energy}) shows that the screened potential (\ref{A1}) is
able to bind particles, a property that is absent in the purely
one-dimensional Coulomb potential.  The result (\ref{energy}) is equivalent to
that obtained in Ref. \cite{Dominguezadame}.

As the potential depth increases for large values of $V_0$, the energy
eigenvalue of any bound state also decreases. When this energy reaches $E=-m$%
, the bound states merge with the negative energy continuum and the
potential becomes supercritical. When $E=-m,$ $\mu =0$ and we obtain from
Eq. (\ref{energy}) the condition for supercriticality: 

\begin{figure}
\includegraphics[width=12cm]{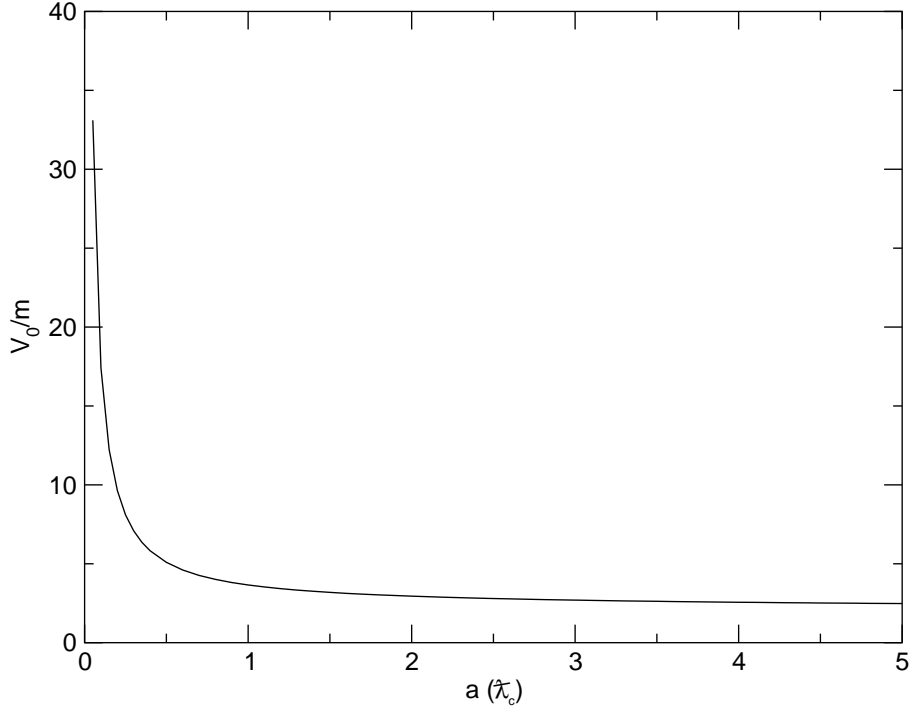}
\caption{\label{fig:Fig2} Dependence of the
supercritical potential amplitude $V_0$ on the potential shape
$a(\lambdabar_{c})$.}
\end{figure}

\begin{equation}
M_{k,0}(2ieaV_0)M_{k,0}(2ieaV_0)-M_{k+1,0}(2ieaV_0)M_{k+1,0}(2ieaV_0)=0
\label{super}
\end{equation}
The condition (\ref{super}) permits one to compute the value of $V_0$ for a
given shape parameter $a$. Fig.~\ref{fig:Fig2}. shows how depends the potential depth on
the shape parameter $a.$

\section{Scattering states}

In order to discuss the transmission and reflection across the barrier given
by the potential (\ref{A}) , for $V_0>0,$we identify the incoming traveling
wave. Since the vector potential (\ref{A1}) vanishes as $x\rightarrow
-\infty $, \ we have the result that the incoming wave should resemble a plane wave for
large negative values of $x$. Looking at the asymptotic behavior of the
Whittaker function $M_{k,\mu }(z)$ as $z\rightarrow 0$ 
\begin{equation}
M_{k,\mu }(z)\rightarrow \exp (-z/2)z^{\frac{1}{2}+\mu },  \label{whi1}
\end{equation}
it is not difficult to obtain that the incoming wave reads 
\begin{equation}
\Psi _{inc}(\bar{y})=\left( 
\begin{array}{c}
\bar{y}^{-1/2}M_{k,\mu }(\bar{y}) \\ 
-\frac{\frac{1}{2}+\mu +k}{ma}\bar{y}^{-1/2}M_{k+1,\mu }(\bar{y})
\end{array}
\right)   \label{inco}
\end{equation}
With the help of the relation (\ref{whi1}), we obtain that, as $x\rightarrow
-\infty ,$ the spinor $\Psi _{inc}$ takes the form 
\begin{equation}
\Psi _{inc}\rightarrow \left( 
\begin{array}{c}
1 \\ 
-\frac{\frac{1}{2}+\mu +k}{ma}
\end{array}
\right) (2ieaV_0)^{\mu }\exp (ix\sqrt{E^{2}-m^{2}})  \label{as1}
\end{equation}
and consequently the spinor (\ref{inco}) satisfies the required asymptotic
behavior. Analogously, we have that the reflected wave can be written as 
\begin{equation}
\Psi _{refl}(\bar{y})=\left( 
\begin{array}{c}
\bar{y}^{-1/2}M_{k,-\mu }(\bar{y}) \\ 
-\frac{\frac{1}{2}-\mu +k}{ma}\bar{y}^{-1/2}M_{k+1,-\mu }(\bar{y})
\end{array}
\right)   \label{ref}
\end{equation}
as $x\rightarrow -\infty ,$ we have that the reflected wave $\Psi _{refl}$
has the asymptotic \ form 
\begin{equation}
\Psi _{refl}\rightarrow \left( 
\begin{array}{c}
1 \\ 
-\frac{\frac{1}{2}-\mu +k}{ma}
\end{array}
\right) (2ieaV_0)^{-\mu }\exp (-ix\sqrt{E^{2}-m^{2}})
\end{equation}
The transmitted wave $\Psi _{trans}$ has the form 
\begin{equation}
\Psi _{trans}(y)=\left( 
\begin{array}{c}
\frac{\frac{1}{2}-\mu +k}{ma}y^{-1/2}M_{k+1,-\mu }(y) \\ 
y^{-1/2}M_{k,-\mu }(y)
\end{array}
\right)   \label{trans}
\end{equation}
whose asymptotic behavior for large values of $x$ is 
\begin{equation}
\Psi _{trans}\rightarrow \left( 
\begin{array}{c}
\frac{\frac{1}{2}-\mu +k}{ma} \\ 
1
\end{array}
\right) (2ieaV_0)^{\mu }\exp (ix\sqrt{E^{2}-m^{2}})
\end{equation}
The computation of the reflection and transmission coefficients can be
achieved with the help of the continuity of the spinor wave-function at the
boundary $x=0$. 
\begin{equation}
A\Psi _{inc}(x=0)+B\Psi _{refl}(x=0)=C\Psi _{trans}(x=0)  \label{borde}
\end{equation}
Substituting (\ref{inco}), (\ref{ref}) and \ (\ref{trans}) into (\ref{borde}%
), we obtain 
\begin{equation}
\frac{B}{A}=\frac{M_{k,\mu }(2ieaV_0)M_{k,-\mu }(2ieaV_0)-M_{k+1,\mu
}(2ieaV_0)M_{k+1,-\mu }(2ieaV_0)}{-\frac{(\frac{1}{2}-\mu +k)^{2}}{m^{2}a^{2}}%
M_{k+1,-\mu }^2(2ieaV_0)-M_{k,-\mu }^2(2ieaV_0)}
\label{res}
\end{equation}
The condition $B/A=0$ is equivalent to considering that there is no reflected
wave, i.e, a transmission resonance condition. From Eq. (\ref{res}) we obtain
that a half bound state or zero momentum resonance condition $\mu =0$
satisfies the equation
\begin{widetext}
\begin{equation}
M_{k,0}(2ieaV_0)M_{k,0}(2ieaV_0)-M_{k+1,0}(2ieaV_0)M_{k+1,0}(2ieaV_0)=0  \label{reso}
\end{equation}
\end{widetext}
Equation (\ref{reso}) gives the values of $V_0$ \ and $a$ for which zero momentum
resonances exist. This condition, with $E=m,$ and $V_0>0,$ is
equivalent to (\ref{super}) for $E=-m$ and $%
V<0.$ In fact, \ after taking the complex conjugate of Eq. (\ref{reso}), \
using the the relation \cite{Gradshtein}   
\begin{equation}
z^{-\frac{1}{2}}M_{k,0}(z)=(-z)^{-\frac{1}{2}}M_{-k,0}(-z)
\end{equation}
for the Whittaker functions, and substituting the value of  $k$ 
defined by the relation (\ref{ka}), we reobtain the supercriticality
condition (\ref{super}). The dependence of $V_0$ on $a$ for transmission 
resonance states is shown in Fig. \ref{fig:Fig2}.

\section{Concluding remarks}

The relations (\ref{super}) and (\ref{reso}) show that the one dimensional
screened potential (\ref{A1}) supports supercritical states and consequently
half bound states. This result is not trivial in view of the fact that the
one-dimensional Coulomb potential presents only scattering states \cite{Dominguez2}. 
The potential (\ref{A1}) does not exhibit a square barrier limit, and for very small 
values of $a$ and constant value of $V_{0}a$ it can be regarded as a delta potential 
\cite{Dominguezadame}; therefore the relations (\ref{super}) and (\ref{reso}) 
also hold for a delta potential. The results obtained in this paper show that very
cusp symmetric potentials also support half bound states. 

. 

\begin{acknowledgments}

One of the author (VMV) wants to express his gratitude to the Alexander von Humboldt
Foundation for financial support.

\end{acknowledgments}


\begin{thebibliography}{99}

\bibitem{Newton}  R. G. Newton Scattering Theory of Waves and Particles
(Springer, Berlin, 1982)

\bibitem{Bohm}  D. Bohm Quantum Mechanics (Prentice-Hall New York, 1951)

\bibitem{Senn}  P. Senn, Am. J. Phys. \textbf{56}, 916 (1988).

\bibitem{Sassoli}  M. Sassoli de Bianchi, J. Math. Phys. \textbf{35}, 2719
(1994).   

\bibitem{Kennedy2} P. Kennedy and N. Dombey,  J. Phys. A \textbf{35}, 6645
(2002) 

\bibitem{Dombey}  N. Dombey, P. Kennedy, and A. Calogeracos Phys. Rev. Lett. 
\textbf{85} 1787 (2000).

\bibitem{Dirac} P. A. M. Dirac, Proc. R. Soc. A \textbf{117}, 610 (1928).

\bibitem{Klein}  O. Klein, Z. Phys. \textbf{53}, 157 (1929).

\bibitem{Sauter}  N. Sauter, Z. Phys. \textbf{69}, 742 (1931).

\bibitem{Greiner}  W. Greiner, B. M\"{u}ller, and J. Rafelski  Quantum
Electrodynamics of Strong Fields (Springer: Berlin, 1985)

\bibitem{Dosch}  H. G. Dosch, J. H. D. Jensen, and V. F. M\"{u}ller, Phys.
Norvegica \textbf{5,} 151 (1971).

\bibitem{Kennedy}  P. Kennedy, J. Phys. A \textbf{35,} 689 (2002).

\bibitem{Abramowitz}  M. Abramowitz, and I. A. Stegun, Handbook of
Mathematical Functions (New York: Dover, 1965).

\bibitem{Dominguezadame}  F. Dom\'{\i}nguez-Adame, and A. Rodr\'{\i}guez, 
Phys. Lett. A. \textbf{198}, 275 (1995).

\bibitem{Gradshtein}  I. S. Gradshtein, and I. W. Ryzhik, Table of
Integrals, Series and Products (Academic, New York, 1984).

\bibitem{Dominguez2} F. Dom\'{\i}nguez-Adame, and E. Maci\'a, Europhys.
Lett. \textbf{8}, 711 (1989). 

\end{thebibliography}
\end{document}